\documentclass[preprint]{elsart}
\usepackage{graphicx}
\begin{document}

%
% !!! needed for "draftcopy"
%

\begin{frontmatter}
\title{Can Experiments Studying Ultrahigh Energy Cosmic Rays Measure
the Evolution of the Sources?  }

\author[Rutgers]{D.R.~Bergman}
\author[Rutgers]{G.B.~Thomson}
\author[Utah]{T.~Abu-Zayyad}
\author[Rutgers]{G.~Hughes}
\author[Utah]{C.C.H.~Jui}
\author[Utah]{J.N.~Matthews}
\author[Utah]{P.~Sokolsky}
\author[Utah,Hawaii]{B.T.~Stokes}
\author[Rutgers,LPNHE]{A.~Zech}

\address[Rutgers]{Rutgers - The State University of New Jersey,
  Department of Physics and Astronomy, Piscataway, New Jersey, USA} 
\address[Utah]{University of Utah, Department of Physics and High
  Energy Astrophysics Institute, Salt Lake City, Utah, USA} 
\address[Hawaii]{University of Hawaii, Department of Physics,
  Honolulu, Hawaii, USA}  
\address[LPNHE]{LPNHE, University of Paris, Paris, France}

\date{\today}

\begin{abstract}

Interactions between cosmic ray protons and the photons of the cosmic
microwave background radiation, as well as the expansion of the
universe, cause cosmic rays to lose energy in a way that depends on
the distance from the cosmic nray source to the earth.  Because of
this, there is a correlation between cosmic ray energies and the
average redshift of their origin.  This correlation may be exploited
to measure the evolution of the sources of cosmic rays.

Sky surveys of Quasi Stellar Objects (QSO's) and Active Galactic
Nuclei (AGN's), made at optical and x-ray wavelengths, are consistent
in showing that the evolution of such objects exhibits a break at a
redshift, $z$, of about 1.6.  At smaller redshifts, the luminosity
density of QSO's and AGN's follows a $(1+z)^m$ distribution, with $m
\sim 2.6$, and exhibit a much flatter distribution above the break.
Measurements of the star formation rate are also consistent with this
picture.

If QSO's and AGN's are sources of ultrahigh energy cosmic rays the
break in their evolution should appear in the cosmic ray spectrum at
an energy of about $10^{17.6}$ eV.  This is the energy of the second
knee.

\end{abstract}
\end{frontmatter}

%\setcounter{page}{1}
%\vspace{0.2in}

\section{Introduction}

Some of the most interesting questions in physics today involve 
ultrahigh energy cosmic rays:  what is their origin and how do they
interact with the Cosmic Microwave Background Radiation (CMBR)
\cite{turner}.  Identifying sources by searching for anisotropy is
made difficult by the fact that the cosmic rays are charged and 
galactic and extragalactic magnetic fields have a strong effect on
their trajectories.

A second technique for studying the sources of cosmic rays uses
spectrum \cite{ds123} and composition \cite{gregpaper} measurements.
These measurements are detailed enough that one can identify, in a
plausible way, the flux both of galactic and extragalactic cosmic
rays.  To learn about the sources of the extragalactic cosmic rays,
one can make a model of these sources and fit it to the data
\cite{fitting}.  This process is aided by the fact that recent
composition measurements indicate that the transition from galactic
sources of cosmic rays to extragalactic sources is complete by about
$10^{18}$ eV, and that throughout the $10^{17}$ eV decade a
considerable fraction of cosmic rays are of extragalactic origin.

There are three energy loss mechanisms that affect extragalactic
cosmic ray protons.  Two of these mechanisms are interactions with
photons of the CMBR: pion production, which causes the GZK suppression
\cite{gzk}, and $e^+e^-$ pair production \cite{Berezinsky}.  The third
mechanism is the expansion of the universe.  These energy loss
mechanisms cause there to be a correlation between the energy of
cosmic rays and the average redshift, $z$, of their sources; i.e.,
cosmic ray protons of a given energy come, on average, from sources at
a certain redshift.  This correlation can be exploited to measure the
evolution of cosmic ray sources.

Astronomical surveys of the distance and luminosity of QSO's and AGN's
have been performed at optical and x-ray wavelengths \cite{qsoagn}.
Upon correcting for observational biases, the luminosity density of
QSO's and AGN's has been measured.  These measurements show that the
$z$-dependence follows a $(1+z)^m$ distribution, with $m \sim 2.6$,
for redshifts less than about 1.6, and exhibit a much flatter
distribution for higher redshifts.  This picture is also consistent
with the star formation rate as measured at infrared wavelengths
\cite{irsfr}.

An interpretation of these observations is that at earlier times, or
at higher redshifts, the large black holes that power the sources were
just forming.  By a redshift of about 1.6 they reached their mature
state, and the source density subsequently follows an evolution
similar to that of the universal expansion.

This means that, if QSO's and AGN's are sources of cosmic rays, their
luminosity at earlier times was considerably lower than one would
expect from observations of closer sources.  Because of the
correlation between cosmic ray energy and the redshift of their
sources, the break in the source luminosity density should show up as
a break in the spectrum of extragalactic cosmic rays, with lower
fluxes at lower energies than one would expect if the source evolution
were smooth.  A break in the source evolution at z=1.6 should produce
a break in the spectrum at an energy of about $10^{17.6}$ eV.  Near
this energy there is a feature in the cosmic ray spectrum called the
second knee.

In this paper, we first describe the spectrum and composition
measurements of the High Resolution Fly's Eye (HiRes) experiment.
Next we describe a model of galactic and extragalactic cosmic rays
that fits both measurements simultaneously.  We then introduce the
evolution of QSO's and AGN's into the model to see the effect in the
region of the second knee.  Finally, we comment on the further
experimental work that must be done to learn more about the sources of
extragalactic cosmic rays.

\section{HiRes Spectrum and Composition Measurements}

\begin{figure}
  \begin{center}
  \includegraphics[width=3in]{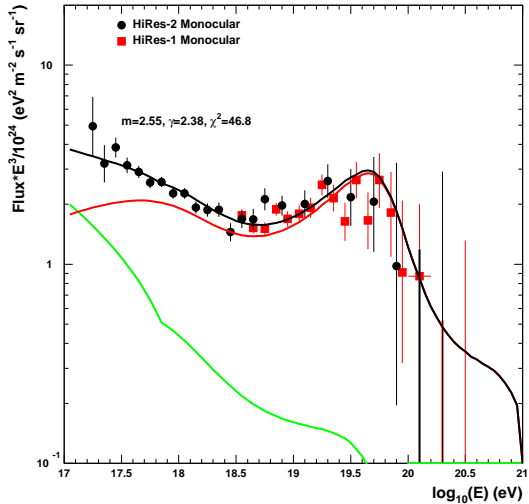}
  \caption{Spectrum of ultrahigh energy cosmic rays as measured by the
    HiRes experiment.  Also shown is the result of a fit to the data
    of a model incorporating galactic (green) and extragalactic (red)
    sources (with the sum in black) described in the text.}
  \label{fig:spect}
  \end{center}
\end{figure}

\begin{figure}
  \begin{center}
  \includegraphics[width=3in]{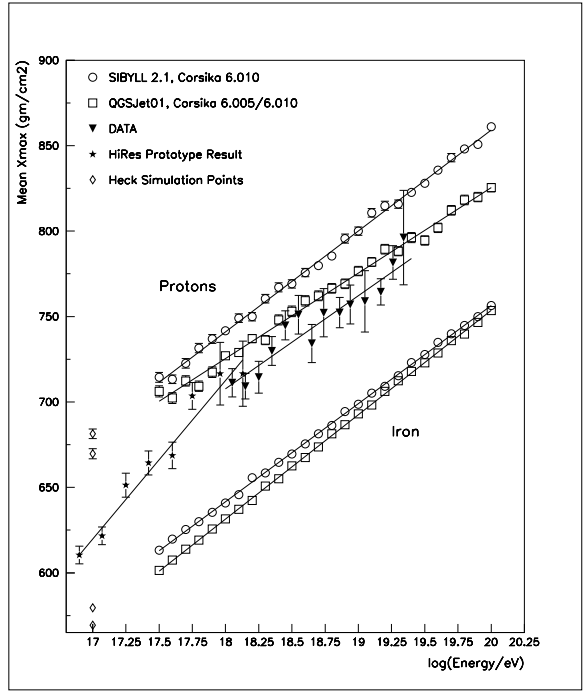}
  \caption{Mean Depth of Shower Maximum, X$_{max}$, measured by the
    HiRes/MIA and HiRes Stereo experiments.  The data, plus
    predictions of Corsika and QGSjet or Sibyll are shown.  The plot
    indicates a transition from heavy to light nuclei (the galactic -
    extragalactic transition) from $10^{17}$ to $10^{18}$ eV.  The
    transition is complete by about $10^{18}$ eV.  The composition is
    light and constant above this energy.}
  \label{fig:comp}
  \end{center}
\end{figure}

The measurements of the cosmic ray spectrum \cite{ds123} and composition
\cite{gregpaper} by the HiRes experiment are shown in Figures
\ref{fig:spect} and \ref{fig:comp}.  The main features of the spectrum
are a strong dip near $10^{18.5}$ eV called the ``ankle'', and a
suppression above $10^{19.8}$ eV which is consistent with the GZK
cutoff.  Below about $10^{17.5}$ eV, the HiRes data have large
statistical (and systematic) uncertainties, and one cannot claim to
see the second knee in these data.

Figure \ref{fig:comp} shows the mean value of
$X_{max}$, the slant depth of shower maximum, as a function of
$\log{E}$.  Results from the HiRes Prototype - MIA hybrid experiment
are shown at lower energies, and of HiRes stereo above $10^{18}$ eV.
The line fit to the HiRes-MIA data has a slope of 93 g/cm$^2$/decade,
where at higher energies the slope is 55 g/cm$^2$/decade.  For
comparison the results of Corsika and two hadronic generator programs,
QGSjet and Sibyll, are shown.  An interpretation of these data is that
below $10^{18}$ eV the composition is getting lighter, and above it is
constant; i.e., since the highest energy galactic cosmic rays are
expected to have heavy composition, and extragalactic to be mostly
protons, the transition from galactic to extragalactic cosmic rays is
complete by about $10^{18}$ eV.

\vspace{0.5in}

\section{A Model of Galactic and Extragalactic Cosmic Rays}

One can construct a model that agrees with all the data described in
the previous section.  If one uses the QGSjet prediction for mean
$X_{max}$ as a guide, the fraction of cosmic rays that are protons is
about 50\% at $10^{17}$ eV and about 80\% for energies above $10^{18}$
eV.  In this model, we ascribe the protons to extragalactic sources and
use the three energy loss mechanisms described above in tracing their
path from source to detection.

In Figure \ref{fig:spect}, the curve drawn through the data is the
result of this model.  We assume the galactic/extragalactic mixture
described above, that the spectrum at the source is a power law of
index $\gamma$ which continues to $10^{21}$ eV then is cut off
sharply.  The source density is a constant times a factor $(1+z)^m$,
where $m$ is known as the evolution parameter.  In this fit, one value
of $m$ is allowed for all redshifts.  The values of $m$ and $\gamma$
are allowed to vary in the fit.  A complete calculation of energy
losses is performed, and pion production is treated by the Monte Carlo
method.

\begin{figure}
  \begin{center}
  \includegraphics[width=4in]{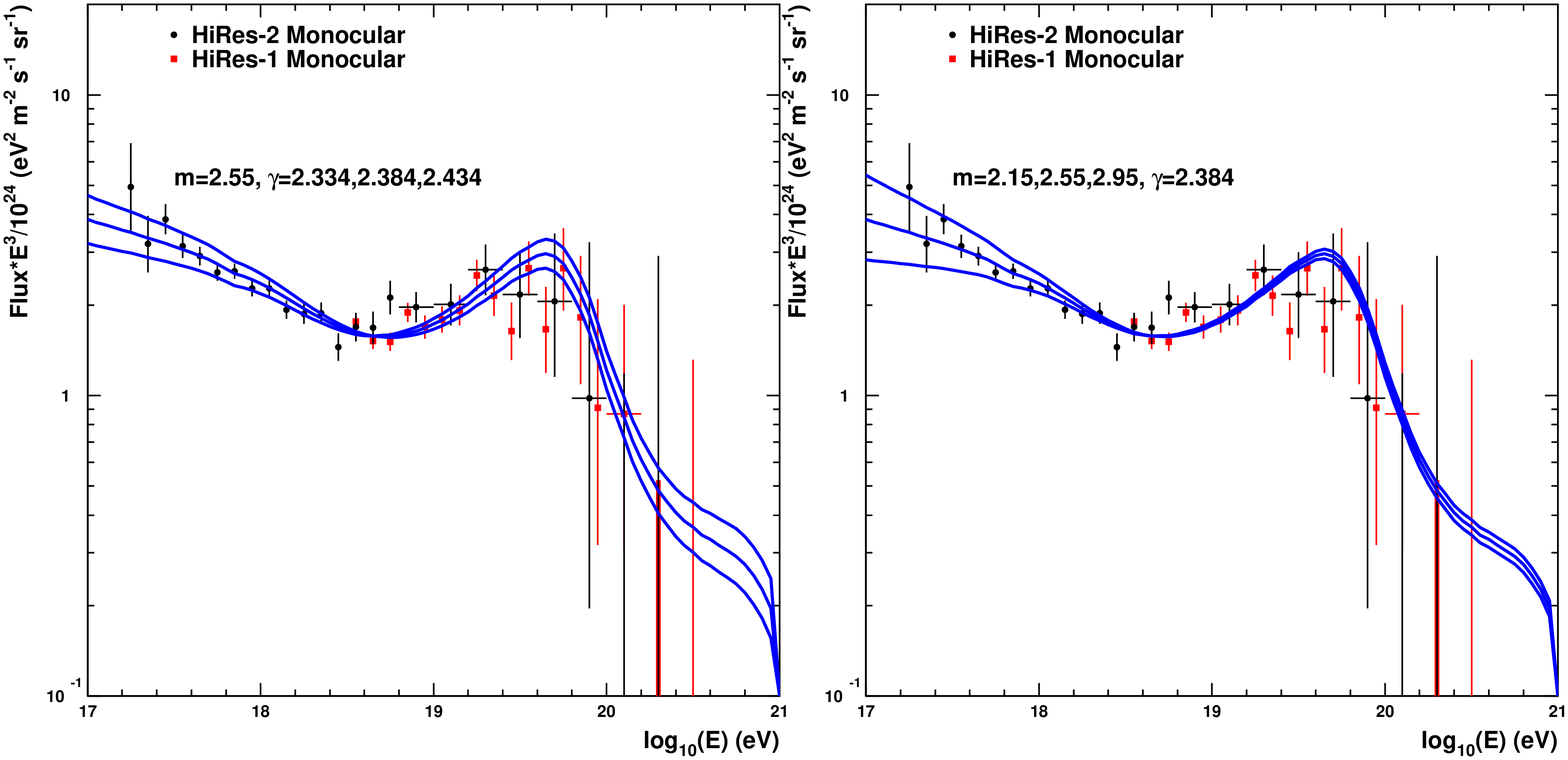}
  \caption{Effect of changing the spectral index, $\gamma$ (left), and
    evolution parameter, $m$ (right), in fits to the HiRes spectrum,
    showing that the ankle region is sensitive to $\gamma$, and that
    the region just below the ankle is most sensitive to m.}
  \label{fig:chisquared}
  \end{center}
\end{figure}

The fit has a satisfactory $\chi^2$ of 43 for 35 degrees of freedom.
It agrees well with the data in the region of the ankle.  The ankle is
especially important because, in this region, the fit is sensitive to
both the average power law index and to the evolution parameter.  The
two parts of Figure \ref{fig:chisquared} show the same data with three
fits in each part, with different values of $\gamma$ ($m$) on the
right (left) side.  The spectrum is sensitive to $\gamma$ throughout
the ankle region, and $m$ on the lower side of the ankle, showing that
the two parameters' values can be extracted from the data.

Between $10^{19.4}$ and $10^{19.7}$ eV the fit is above the
data points.  This may be due to the model's abrupt cutoff at
$10^{21}$ eV, whereas sources really have a distribution of maximum
energies.  This is an interesting piece of information that bears on
the distribution of the size and magnetic fields of sources, and
should be investigated further.

Experiments studying lower energy regions than HiRes \cite{akeno}
observe a spectrum that is flat on an $E^3J$ plot (i.e., falls like
$E^{-3.0}$) below the second knee.  The second knee was observed by
the Fly's Eye experiment at $10^{17.6}$ eV.  The behavior of the model
below $10^{17.6}$ eV is not what one would expect from the world's
data: the model's prediction is that the spectrum should be steeper,
at about $E^{-3.2}$.  Another way of saying this is that the
appearance of the second knee is too weak in this model.

\begin{figure}
  \begin{center}
  \includegraphics[width=4in]{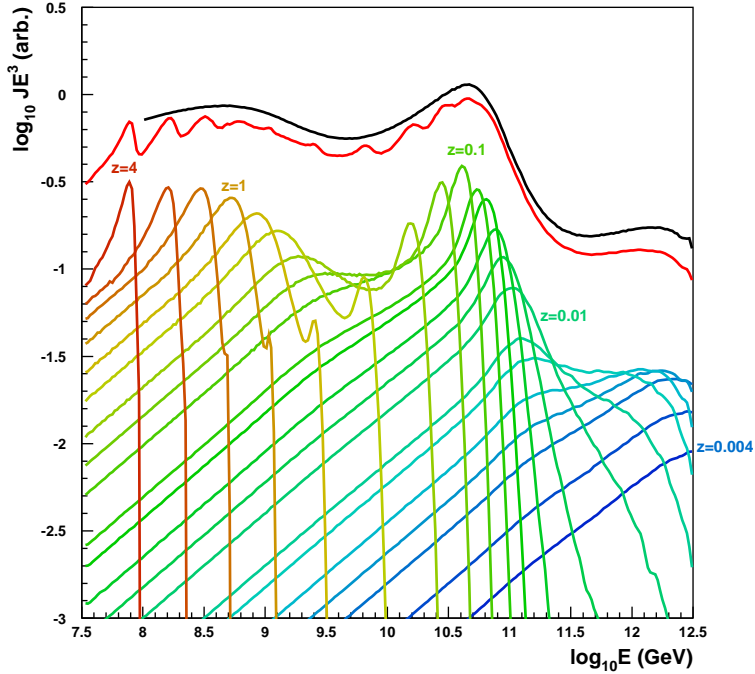}
  \caption{Contributions from sources within shells in redshift
    according to our energy loss model.}
  \label{fig:shells}
  \end{center}
\end{figure}

Figure \ref{fig:shells} shows how sources grouped in shells of
redshift contribute to the cosmic ray spectrum, and the correlation
between energy and redshift.  As an example, one can see the
correlation by observing that at $\log{E}$ of 8.8 ($E$ in GeV), at the
peak of the shell at z=1.0, the contribution from the shell at z=0.1
is lower at this energy by an order of magnitude.  The correlation is
not perfect, but it is significant.

This correlation is due to the effect of energy loss mechanisms
working on cosmic rays as they traverse large distances across the
universe.  To again take the example of sources at z=1.0, the highest
energy cosmic rays, in this model $10^{21}$ eV, are reduced to an
energy of about $10^{17.9}$ eV at the earth by these effects.

\section{QSO and AGN Evolution and the Second Knee}

Astronomical sky surveys have been performed for QSO's and AGN's at
optical and X-ray wavelengths \cite{qsoagn}.  In such a survey,
sources are tabulated by magnitude and redshift.  After correcting for
observational biases, the luminosity density of these astronomical
objects is normally calculated as a function of redshift.  Surveys at
each wavelength consistently show a density that rises at low $z$ like
$(1+z)^m$, with $m \sim 2.6$.  At higher $z$, the density flattens out
considerably.  Infrared measurements of the star formation rate
\cite{irsfr} are consistent with this picture.

The interpretation is that as time progressed - or as $z$ decreased -
the large black holes that power QSO's and AGN's first grew to
significant size, then their activity reached a plateau at $z \sim
1.6$.  The source evolution after this almost followed that of
the universal expansion (which would be m=3).  The effect of this, on
the flux of extragalactic cosmic rays, is that at energies lower than
about $10^{17.6}$ there would be fewer cosmic rays than one would
expect from the density of sources at redshifts less than 1.6.

\begin{figure}
  \begin{center}
    \begin{minipage}[t]{0.49\columnwidth}
      \includegraphics[width=\columnwidth]{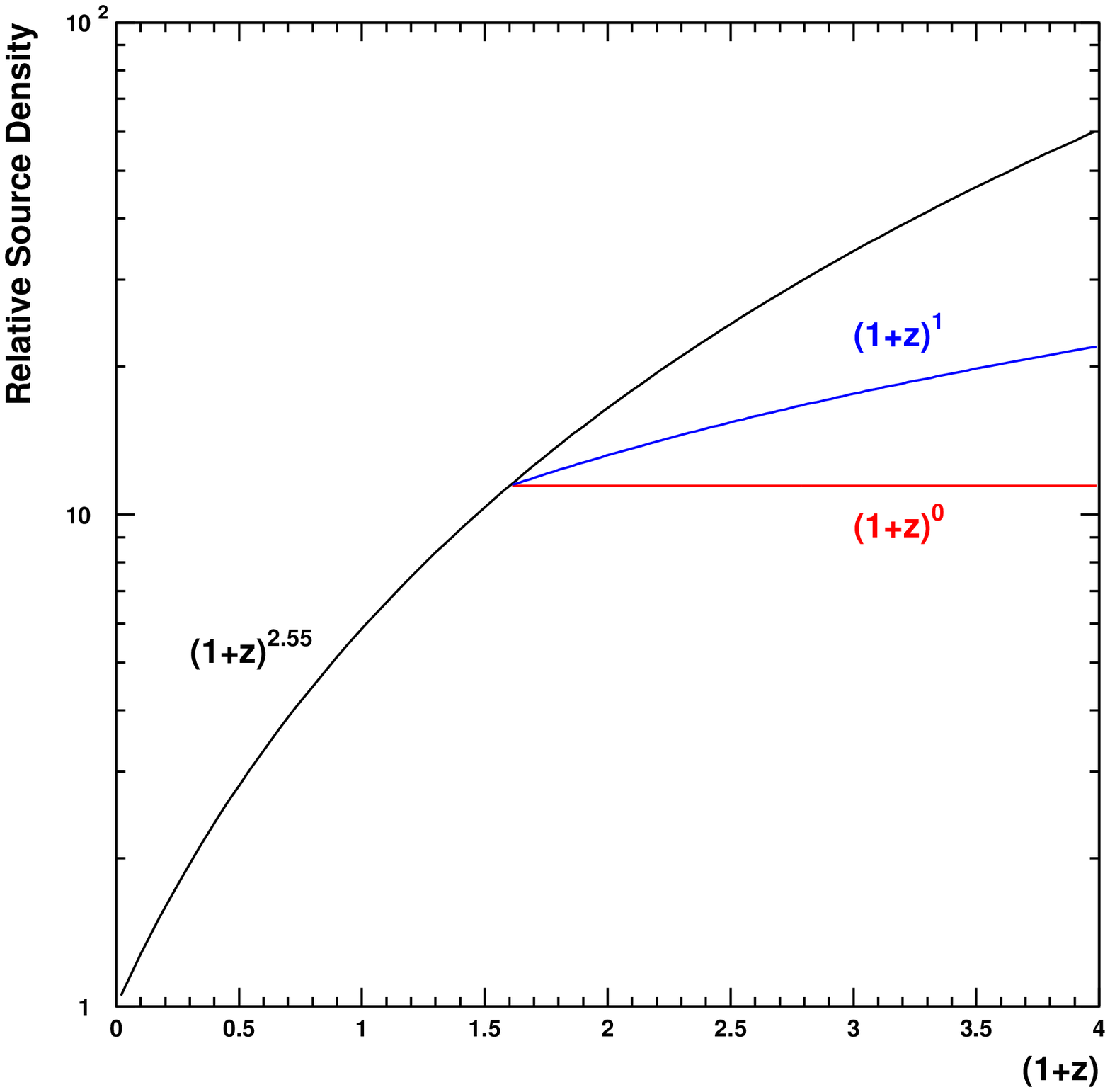}
    \end{minipage}
    \begin{minipage}[t]{0.49\columnwidth}
      \includegraphics[width=\columnwidth]{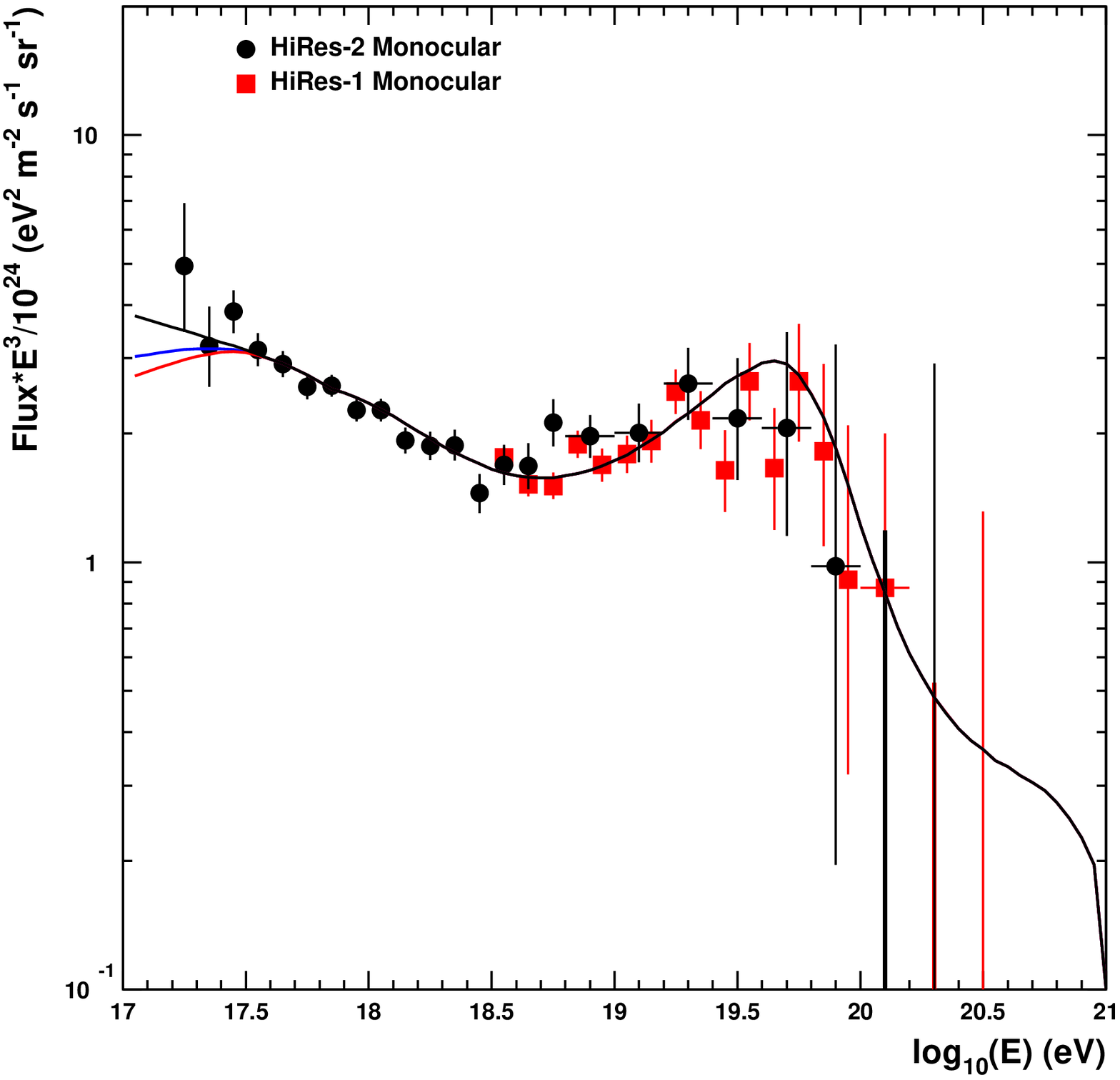}
    \end{minipage}
  \end{center}
  \caption{Left figure: Three curves of the density of cosmic
    ray sources.  Evolution as $(1+z)^{2.6}$ for all $z$ (black),
    $(1+z)^1$ above $z=1.6$ (blue), and $(1+z)^{0}$ above $z=1.6$
    (red).  Right figure: Model fits to the HiRes data using the three forms
    of source evolution as in the left part of this figure.  Note the
    variation below $10^{17.5}$ eV.  This indicates that a more
    precise measurement of the spectrum below $10^{17.5}$ eV can be
    sensitive to the evolution of the sources.}
  \label{fig:knee2}
\end{figure}

To test the effect of the break in source evolution on the cosmic ray
spectrum, we introduced the break into our model of galactic and
extragalactic cosmic rays.  The left part of Figure \ref{fig:knee2}
shows three density curvess as a function of redshift.  In black is
the standard $(1+z)^{2.6}$ as in previous figures, in blue is
$(1+z)^{1}$ above z=1.6, and in red is $(1+z)^{0}$ above z=1.6.  The
result of including these density functions is shown in the right half
of Figure \ref{fig:knee2}.  This shows that the affect of changing the
evolution of cosmic ray sources to conform to the break in QSO and AGN
evolution at z=1.6 also appears as a break in the cosmic ray spectrum
at an energy of $10^{17.6}$ eV.  This is the approximate location of
the ``second knee'' of the cosmic ray spectrum.  The $(1+z)^{1}$
density function yields the flattest spectrum below $10^{17.6}$ eV.

We do not believe that this constitutes a determination of the
evolution of cosmic ray sources, but rather is an example of what
might be done with data from a well-designed future cosmic ray
experiment.  

\section{Future Experiments}

The modeling exercise above clearly shows the limitations of the
present state of the world's cosmic ray data.  Two essential elements
of fits like these are a measurement of the spectrum of cosmic rays by
a single experiment covering a very wide energy range, and a
measurement of $X_{max}$ again covering a very wide energy range.  The
$X_{max}$ measurement must be made by a fluorescence detector, since
only fluorescence detectors actually observe $X_{max}$.  Ground array
experiments must infer $X_{max}$, with their attendant model-dependent
systematic uncertainties.  In order to be sensitive to the evolution
of cosmic ray sources an experiment must observe all the structures in
the spectrum between the second knee and the GZK cutoff.  In a fit to
the spectrum, each structure will yield information about cosmic ray
sources.  Hence it is necessary for a single experiment to have
fluorescence coverage between about $10^{16.5}$ and $10^{20.5}$ eV.

Although the HiRes experiment covers about three orders of magnitude
in energy, from $10^{17.5}$ to $10^{20.5}$ eV, and can measure the
evolution parameter, it does not reach low enough energies to be
sensitive to a break at z=1.6 (an energy of $10^{17.6}$ eV) in the
evolution of QSO's and AGN's via the cosmic ray spectrum.

The sole experiment, built or planned, that has the capability of
performing this test is the Telescope Array (TA).  The TA experiment,
currently being constructed in Millard County, Utah, will consist of a
ground array of 576 scintillation counters, deployed on a grid of
spacing 1.2 km, and overlooked by three fluorescence detectors for
hybrid coverage.  The TA experiment will be fully efficient above
$10^{19}$ eV.

Two further detectors are planned to extend the coverage in stereo and
hybrid mode to lower energies.  The first is a pair of fluorescence
detectors located 6 km from two of the main TA fluorescence detectors.
These will observe the ankle region stereoscopically.  They will
extend the stereo coverage below $10^{18}$ eV and will provide an
excellent measurement of both the average power law of extragalactic
sources and the evolution parameter.  The second is a detector with
larger mirrors observing cosmic ray showers at higher elevation
angles.  It will be deployed with an infill array in front of it and
will cover the energy range from $10^{16.5}$ to $10^{18}$ eV in hybrid
mode.  This detector is designed to study the region of the second
knee.  This suite of detectors is called the Telescope Array Low
Energy Extension (TALE).

The TA and TALE detectors will provide seamless coverage over four
decades in energy, from $10^{16.5}$ to $10^{20.5}$ eV.  The same
events will be seen by several of the detectors and cross correlation
of energy scales will be possible.  Only a suite of detectors like
this can measure the evolution of extragalactic cosmic ray sources.

Performing a study of spectral structure, in correlation with observed
$X_{max}$, can determine if the second knee is of galactic or
extragalactic origin.  One can select events based on deep $X_{max}$
values to isolate the protonic, and hence extragalactic, component of
the cosmic ray flux (and conversely choose the heavy or galactic
component by choosing events with shallow $X_{max}$).  If the second knee
shows up in the extragalactic component, it strengthens the evolution
argument presented here.  If the second knee proves to be of galactic
origin, it would be very interesting, but would invalidate the
evolution-origin hypothesis.

\section{Summary}

The technique of measuring the spectrum and composition over a wide
energy range by a fluorescence detector is a very powerful one for
understanding the sources of ultrahigh energy cosmic rays.  Observing
in monocular or stereo modes for the spectrum measurement, and in
stereo mode for the composition measurement is important.  Hybrid mode
is only marginally better than monocular mode for spectrum measurement
(it is noticably better for composition studies), but nothing
approaches the excellence of stereo, with its ability to make two
determinations of energy and $X_{max}$ and measure the
uncertainties in those quantities on an event-by-event basis.

Astronomical sky survey experiments have observed that the evolution
of QSO's and AGN's exhibits structure at a redshift of $z \sim 1.6$.
If QSO's and AGN's are sources of cosmic rays, this has the implication
that the generation of extragalactic cosmic rays by sources more
distant than z=1.6 is weaker than one would expect from observations
of closer sources.

There is a correlation between the energy of cosmic ray protons and the
redshift of their sources due to the strong energy loss mechanisms at
work when the cosmic rays traverse long distances across the universe.
This means that the break in source evolution should show up as a
break in the extragalactic cosmic ray spectrum as well.  A break at a
redshift of 1.6 should show up at an energy of $10^{17.6}$ eV, which
is the approximate location of the second knee of the cosmic ray
spectrum.

No running experiment has the capability of observing this effect.
The energy range of the HiRes experiment does not extend to low enough
energies for this observation and that of the Auger Observatory is
certainly too narrow.  Only the Telescope Array experiment will cover
a wide enough energy range to test this prediction.

The authors wish to thank M. Gaskell and D. Seckel for suggesting this
topic, and G. Richards and P. Biermann for discussions of the
astronomical survey data.  This work was supported by US NSF grants
PHY-0305516 and PHY-0307098.

\end{document}